\documentclass[a4paper, 11pt]{article}
\usepackage{a4wide}
\usepackage{amsmath}
\usepackage{commath}
\usepackage{latexsym}
\usepackage{graphicx}
\usepackage{color}
\usepackage[pdfstartview=FitH]{hyperref}
\usepackage{verbatim}
\usepackage{authblk}
\usepackage[subtle]{savetrees}

\title{A skew logistic distribution for modelling COVID-19 waves and its evaluation using the empirical survival Jensen-Shannon divergence}

\author{Mark Levene \thanks{corresponding author; email: mlevene@dcs.bbk.ac.uk;  ORCID: 0000-0001-8632-4732}}
\affil{Department of Computer Science and Information Systems, \authorcr Birkbeck, University of London, London WC1E 7HX, U.K.}

\date{}

\begin{document}

\maketitle

\newtheorem{algorithm}{Algorithm}

\begin{abstract}

A novel yet simple extension of the symmetric logistic distribution is proposed by introducing a skewness parameter. It is shown how the three parameters of the ensuing skew logistic distribution may be estimated using maximum likelihood. The skew logistic distribution is then extended to the skew bi-logistic distribution to allow the modelling of multiple waves in epidemic time series data. The proposed skew-logistic model is validated on COVID-19 data from the UK, and is evaluated for goodness-of-fit against the logistic and normal distributions using the recently formulated empirical survival Jensen-Shannon divergence (${\cal E}SJS$) and the Kolmogorov-Smirnov two-sample test statistic ($KS2$). We employ 95\% bootstrap confidence intervals to assess the improvement in goodness-of-fit of the skew logistic distribution over the other distributions. The obtained confidence intervals for the ${\cal E}SJS$ are narrower than those for the $KS2$ on using  this data set, implying that the ${\cal E}SJS$ is more powerful than the $KS2$.

\end{abstract}

\noindent {\it Keywords: }{{empirical survival Jensen-Shannon divergence; Kolmogorov-Smirnov two-sample test; skew logistic distribution; bi-logistic growth; epidemic waves; COVID-19 data.}

\section{Introduction}
\label{sec:intro}

In exponential growth the population grows at a rate proportional to its current size. This is unrealistic, since in reality growth will not exceed some maximum, called its carrying capacity. The logistic equation \cite[Chapter 6]{BACA11} deals with this problem by ensuring that the growth rate of the population decreases once the population reaches its carrying capacity \cite{PANI14}. Statistical modelling of the logistic equation's growth and decay is accomplished with the {\em logistic distribution} \cite{JOHN95c} \cite[Chapter 22]{KRIS15}, noting that the tails of the logistic distribution are heavier than those of the ubiquitous normal distribution. The normal and logistic distributions are both symmetric, however, real data often exhibits skewness \cite{DASG10}, which has given rise to extensions of the normal distribution to accommodate for skewness, as in the skew normal \cite{AZZA14} and epsilon skew normal \cite{MUDH00} distributions. Subsequently, skew logistic distributions were also devised, as in \cite{NADA09a,SAST16}.

\smallskip

Epidemics, such as COVID-19, are traditionally modelled by compartmental models such as the SIR (Susceptible-Infected-Removed) model and its extension the SEIR (Susceptible-Exposed-Infected-Removed) model, which estimate the trajectory of an epidemic \cite{LI18}. These models typically rely on assumptions on how the disease is transmitted and progresses \cite{IOAN22}, and are routinely used to understand the consequences of policies such as mask wearing and social distancing \cite{DAVI20}. Time series models \cite{HARV21}, on the other hand, employ historical data to make forecasts about the future, are generally simpler than compartmental models, and are able to make forecasts on, for example, number of cases, hospitalisations and deaths. The SIR model can be interpreted as a logistic growth model \cite{DELA20,POST20}. However, as the data is inherently skewed, a skewed logistic statistical model would  be a natural choice, although as such  it does not rely on biological assumptions in its forecasts \cite{DYE20}.

\smallskip

Herein we present a novel yet simple  (one may argue the simplest), three parameter skewed extension to the logistic distribution to allow for asymmetry; c.f. \cite{DYE20}. Nevertheless, if instead of our extension we deploy one of the other skew logistic distributions (such as the one described in \cite{NADA09a}) the results, would no doubt be comparable to the results we obtain herein; we, however, pursue our simpler extension detailing its statistical properties.

\smallskip

In the context of analysing epidemics the logistic distribution is normally preferred, as it is a natural distribution to use in modelling population growth and decay. However, we still briefly mention a comparison of the results we obtain in modelling COVID-19 waves with the skew logistic distribution, to one which, instead, employs a skew normal distribution (more specifically we choose the, flexible, epsilon skew normal distribution \cite{MUDH00}). The result of this comparison implies that utilising the epsilon skew normal distribution leads, overall, to results which are comparable to those when utilising the skew logistic distribution. However, in practice, it is still preferable to make use of the skew logistic distribution as it is the natural model to deploy in this context \cite{PELIN20}, since, on the whole, it is more consistent with the data as its tails are heavier than those of a skew normal distribution.

\smallskip

Epidemics are said to come in ``waves''. The precise definition of a wave is somewhat elusive \cite{ZHAN21}, but it is generally accepted that, assuming we have a time series of the number of, say, daily hospitalisations, a wave will span over a period from one valley (minima) in the time series to another valley, with a peak (maxima) in between them; there is no strict requirement that waves do not overlap, although, here for simplicity we will not consider any overlap as such; see \cite{ZHAN21} for an attempt to give an operational definition of the concept of epidemic wave. In order to combine waves we make use of the concept of {\em bi-logistic growth} \cite{MEYE94,FENN13}, or more generally multi-logistic growth, which allows us to sum two or more instances logistic growth when the time series spans over more than a single wave.

\smallskip

To fit the skew logistic distribution to the time series data we employ maximum likelihood, and to evaluate the goodness-of-fit we make use of the recently formulated {\em empirical survival Jensen-Shannon divergence} (${\cal E}SJS$) \cite{LEVE18,LEVE21a} and the well-established {\em Kolmogorov-Smirnov two-sample test statistic} ($KS2$) \cite[Section 6.3]{GIBB21}. The ${\cal E}SJS$ is an information-theoretic goodness-of-fit measure of a fitted parametric continuous distribution, which overcomes the inadequacy of the {\em coefficient of determination}, $R^2$, as a goodness-of-fit measure for nonlinear models  \cite{SPIE10}. The $KS2$ statistic also satisfies this criteria regarding $R^2$, however we observe that the 95\% bootstrap confidence intervals \cite{EFRO93} we obtain for the ${\cal E}SJS$ are narrower than those for the $KS2$, suggesting that the ${\cal E}SJS$ is more powerful \cite{COLE21} than the $KS2$. Another, well-known, limitation of the $KS2$ statistic is that it is less sensitive to discrepancies at the tails of the distribution than the ${\cal E}SJS$ statistic is, in the sense that as opposed to ${\cal E}SJS$ it is ``local'', i.e. its value is determined by a single point \cite{BEND15}.

\smallskip

The rest of the paper is organised as follows.
In Section~\ref{sec:sl}, we introduce a skew logistic distribution, which is a simple extension of the standard, symmetric,  logistic distribution obtained by adding to it a single skew parameter, and  derive some of its properties.
In Section~\ref{sec:ml}, we formulate the solution to the maximum likelihood estimation of the parameters of the skew logistic distribution. In Section~\ref{sec:bilog}, we make use of an extension of the skew logistic distribution to the bi-skew logistic distribution to model a time series of COVID-19 data items having more than a single wave. In Section~\ref{sec:data} we provide analysis of daily COVID-19 deaths in the UK from 30/01/20 to 30/07/21, assuming the skew logistic distribution as an underlying model of the data. The evaluation of goodness-of-fit of the skew logistic distribution to the data makes use of the recently formulated ${\cal E}SJS$, and compares the results to those when employing the $KS2$ instead. We observe that the same technique, which we applied to the analysis of COVID-19 deaths, can be used to model new cases and hospitalisations. Finally, in Section~\ref{sec:conc}, we present our concluding remarks.
It is worth noting that in the more general setting of information modelling, being able to detect epidemic waves may help supply chains in planning increased resistance to such adverse events \cite{SEME22}. We note that all computations were carried out using the Matlab software package.

\section{A skew logistic distribution}
\label{sec:sl}

Here we introduce a novel {\em skew logistic distribution}, which extends, in straightforward manner, the standard two parameter logistic distribution \cite{JOHN95c} \cite[Chapter 22]{KRIS15} by adding to it a skew parameter. The rationale for introducing the distribution is that, apart from its simple formulation, we believe its maximum likelihood solution, presented below is also simpler than those derived for other skew logistic distributions,  such as the ones investigated in \cite{NADA09a,SAST16}. This point provides further justification for our skew logistic distribution when introducing the bi-skew logistic distribution in Section~\ref{sec:bilog}.

\smallskip

Now, let $\mu$ be a location parameter, $s$ be a scale parameter and $\lambda$ be a skew parameter, where $s > 0$ and $0 < \lambda < 2$. Then, the probability density function of the skew logistic distribution at a value $x$ of the random variable $X$, denoted as $f(x;\lambda,\mu,s)$, is given by
\begin{equation}\label{eq:pdf}
f(x;\lambda,\mu,s) = \frac{\kappa_\lambda \ \exp\left(- \lambda \ \frac{x-\mu}{s} \right)}{s \left( 1+ \exp\left(- \frac{x-\mu}{s} \right) \right)^2},
\end{equation}
noting that for clarity we write $x-\mu$ above as a shorthand for $\left( x-\mu \right)$, and $\kappa_\lambda$ is a normalisation constant, which depends on $\lambda$.

\smallskip

When $\lambda = 1$, the~skew logistic distribution reduces to the standard logistic distribution as in~\cite{JOHN95c} and \cite{KRIS15} (Chapter 22), which is symmetric. On~the other hand, when $0 < \lambda < 1$, the~skew logistic distribution is positively skewed, and~when $1 < \lambda < 2$, it is negatively~skewed. So, when $\lambda = 1$, $\kappa_\lambda = 1$, and, for example, when $\lambda = 0.5$ or $1.5$, $\kappa_\lambda = 2/\pi$. For simplicity, from now on, unless necessary, we will omit to mention the constant $\kappa_\lambda$ as it will not effect any of the results.

\smallskip

The {\em skewness} of a random variable $X$ \cite{DASG10,KRIS15}, is defined as
\begin{displaymath}
{\rm E}\left[ \left( \frac{X-\mu}{s} \right)^3 \right],
\end{displaymath}
and thus, assuming for simplicity of exposition (due the linearity of expectations \cite{DASG10}) that $\mu=0$ and $s=1$, the skewness of the skew logistic distribution, denoted by $\gamma(\lambda)$,  is given by
\begin{equation}\label{eq:skew1}
\gamma(\lambda) = \int_{-\infty}^\infty x^3 \ \frac{\exp\left(- \lambda x \right)}{s \left( 1+ \exp\left(- x \right) \right)^2} \ dx.
\end{equation}

First, we will show that letting $\lambda_1 = \lambda$, with $0 < \lambda_1 < 1$, we have $\gamma(\lambda_1) > 0$, that is $f(x;\lambda_1,0,1)$ is positively skewed. We can split the integral in (\ref{eq:skew1}) into two integrals for the negative part from $-\infty$ to $0$ and the positive part from $0$ to $\infty$, noting that when $x=0$, the expression to the right of the integral is equal to $0$. Then, on setting $y=-x$ for the negative part, and $y=x$ for the positive part, the result follows, as by algebraic manipulation it can be shown that
\begin{equation}\label{eq:skew2}
\frac{\exp(- \lambda_1 y)}{\left( 1 + \exp(-y) \right)^2} > \frac{\exp(\lambda_1 y)}{\left( 1 + \exp(y) \right)^2},
\end{equation}
implying that $\gamma(\lambda_1) > 0$ as required.

\smallskip

Second, in a similar fashion to above, on letting $\lambda_2 = \lambda_1 +1 = \lambda$, with $1 < \lambda_2 < 2$, it follows that $\gamma(\lambda_2) < 0$, that is $f(x;\lambda_2,0,1)$ is negatively skewed. In particular, by algebraic manipulation we have that
\begin{equation}\label{eq:skew3}
\frac{\exp \left(- \lambda_2 y \right)}{\left( 1 + \exp(-y) \right)^2} < \frac{\exp \left( \lambda_2 y \right)}{\left( 1 + \exp(y) \right)^2},
\end{equation}
implying that $\gamma(\lambda_2) < 0$ as required.

\medskip

The cumulative distribution function of the skew logistic distribution at a value $x$ of the random variable $X$ is obtained by integrating $f(x;\lambda,\mu,s)$, to obtain $F(x;\mu,s,\lambda)$, which is given by
\begin{align}\label{eq:cdf}
F(x;\lambda,\mu,s) = \kappa_\lambda \ \exp\left( -(\lambda-2) \ \frac{x-\mu}{s} \right) & \left( \frac{1}{\left(1+\exp \left(\frac{x-\mu}{s}\right)\right)} \right. - \nonumber \\
& \left. \ \ \frac{\lambda-1}{\lambda-2} \ {}_2F_1\left(1,2-\lambda;3-\lambda;-\exp\left(\frac{x-\mu}{s}\right)\right)\right),
\end{align}
where ${}_2F_1(a,b;c;z)$ is the {\em Gauss hypergemoetric function} \cite[Chapter 15]{ABRA72}; we assume $a, b$ and $c$ are positive real numbers, and that $z$ is a real number extended outside the unit disk by analytic continuation \cite{PEAR17}.

\smallskip

The hypergeometric function has the following integral representation \cite[Chapter 15]{ABRA72},
\begin{equation}\label{eq:hyperg1}
\frac{\Gamma(c)}{\Gamma(b) \Gamma(c-b)} \int_0^1 \frac{t^{b-1} (1-t)^{c-b-1}}{(1 - tz)^a} \ dt,
\end{equation}
where $c > b$. Now, assuming without loss of generality that $u=0$ and $s=1$, we have that
\begin{equation}\label{eq:hyperg2}
{}_2F_1\left(1,2-\lambda;3-\lambda;-\exp (x) \right) = \left( 2-\lambda \right) \int_0^1 \frac{t^{1-\lambda}}{\left( 1 + t \ \exp(x) \right)} \ dt,
\end{equation}
where $x$ is a real number.

\smallskip

Therefore, from (\ref{eq:hyperg2}) it can be verified that: (i) ${}_2F_1 (1,2-\lambda;3-\lambda;-\exp (x))$ is monotonically decreasing with $x$, (ii) as $x$ tends to plus infinity, ${}_2F_1 (1,2-\lambda;3-\lambda;-\exp (x))$ tends to $0$, and (iii) as $x$ tends to minus infinity, ${}_2F_1 (1,2-\lambda;3-\lambda;-\exp (x))$ tends to $1$, since
\begin{displaymath}
\left( 2 - \lambda \right) \int_0^1 t^{1- \lambda} \ dt  = 1.
\end{displaymath}

\section{Maximum likelihood estimation for the skew logistic distribution}
\label{sec:ml}

We now formulate the maximum likelihood estimation \cite{WARD18} of the parameters $\mu, s$ and $\lambda$ of the skew logistic distribution. Let $\{x_1, x_2, \ldots, x_n\}$ be a random sample of $n$ values from the density function of the skew logistic distribution in (\ref{eq:pdf}).  Then, the log likelihood function of its three parameters is given by
\begin{equation}\label{eq:mle}
\ln L(\lambda,\mu,s) = - n \ln(s) - \frac{\lambda}{s} \sum_{i=1}^n (x_i - \mu) - 2 \sum_{i=1}^n  \ln \left( 1+ \exp \left( - \frac{x_i - \mu}{s} \right) \right).
\end{equation}
\smallskip

In order to solve the log likelihood function, we first partially differentiate $\ln L(\lambda,\mu,s)$ as follows:
\begin{align}\label{eq:mle-eqs}
\frac{\partial \ln L(\lambda,\mu,s)}{\partial \lambda} &= \sum_{i=1}^n \frac{\mu - x_i}{s}, \nonumber \\
\frac{\partial \ln L(\lambda,\mu,s)}{\partial \mu}     &= \frac{\lambda n}{s} - \frac{2}{s} \sum_{i=1}^n \frac{1}{1 + \exp \left( \frac{x_i - \mu}{s} \right)} \ {\rm and} \nonumber \\
\frac{\partial \ln L(\lambda,\mu,s)}{\partial s}       &= - \frac{n}{s} + \frac{1}{s^2} \sum_{i=1}^n \left( x_i - \mu \right)
\left( \lambda -  \frac{2}{1 + \exp \left( \frac{x_i - \mu}{s} \right)} \right).
\end{align}
\smallskip

It is therefore implied that the maximum likelihood estimators are the solutions to the following three equations:
\begin{align}\label{eq:mle-sol}
\mu     &= \frac{\sum_{i=1}^n x_i}{n}, \nonumber \\
\lambda &= \frac{2}{n} \sum_{i=1}^n \frac{1}{1 + \exp \left( \frac{x_i - \mu}{s} \right)} \ {\rm and} \nonumber \\
s       &=  \frac{1}{n} \sum_{i=1}^n \left( x_i - \mu \right) \left( \lambda -  \frac{2}{1 + \exp \left( \frac{x_i - \mu}{s} \right)} \right),
\end{align}
which can be solved numerically.
\smallskip


We observe that the equation for $\mu$ in (\ref{eq:mle-sol}) does not contribute to solving the maximum likelihood, since the location parameter $\mu$ is equal to the mean only when $\lambda=1$. We thus look at an alternative equation for $\mu$, which involves the mode of the skew logistic distribution.

\smallskip

To derive the mode of the skew logistic distribution we solve the equation,
\begin{equation}\label{eq:mode1}
\frac{\partial}{\partial x} \frac{\exp\left(- \lambda \ \frac{x-\mu}{s}\right)}{s \left( 1+ \exp\left(- \frac{x-\mu}{s} \right) \right)^2} = 0,
\end{equation}
to obtain
\begin{equation}\label{eq:mode2}
\mu = x - s \ \log \left( - \frac{\lambda-2}{\lambda} \right).
\end{equation}
\smallskip

Thus, motivated by ({\ref{eq:mode2}) we replace the equation for $\mu$ in (\ref{eq:mle-sol}) with
\begin{equation}\label{eq:mode3}
\mu = m - s \ \log \left( - \frac{\lambda-2}{\lambda} \right),
\end{equation}
where $m$ is the mode of the random sample.

\section{The bi-skew logistic distribution for modelling epidemic waves}
\label{sec:bilog}

We start by defining the bi-skew logistic distribution, which will enable us to model more than one wave of infections at a time. We then discuss how we partition the data into single waves, in a way that we can apply the maximum likelihood from the previous section to the data in a consistent manner.

\smallskip

We present the {\em bi-skew logistic distribution}, which is described by the sum,
\begin{displaymath}
f(x;\lambda_1,\mu_1,s_1) + f(x;\lambda_2,\mu_2,s_2),
\end{displaymath}
of two skew logistic distributions. It is given in full as
\begin{equation}\label{eq:bi-skew}
\frac{\exp \left( -\lambda_1 \frac{x - \mu_1}{s_1} \right)}{s_1 \left( 1+ \exp \left(- \frac{x - \mu_1}{s_1} \right) \right)^2} + \frac{\exp \left( -\lambda_2 \frac{x - \mu_2}{s_2} \right)}{s_2 \left( 1+ \exp \left(- \frac{x - \mu_2}{s_2} \right) \right)^2},
\end{equation}
which characterises two distinct phases of logistic growth (c.f. \cite{MEYE94,SHEE04}). We note that (\ref{eq:bi-skew}) can be readily extended to the general case of the sum of multiple skew logistic distributions, however for simplicity we only present the formula for the bi-skew logistic case. Thus while the (single) skew logistic distribution can only model one wave of infected cases (or deaths, or hospitalisations) the bi-skew logistic distribution can model two waves of infections, and in the general cases any number of waves.

\smallskip

In the presence of two waves the maximum likelihood solution to (\ref{eq:bi-skew}) would give us access to the necessary model parameters, and solving the general case in presence of multiple waves, when the sum in (\ref{eq:bi-skew}) may have two or more skew logistic distributions, is evidently even more challenging. Thus we simplify the solution for the multiple wave case, and concentrate on an approximation assuming a sequential time series when one wave strictly follows the next. More specifically, we assume that each wave is modelled by a single skewed logistic distribution describing the growth phase until a peak is reached, followed by a decline phase; see \cite{CLIFF82} who considers epidemic waves in the context of the standard logistic distribution. Thus a wave is represented by a temporal pattern of growth and decline, and the time series as whole describes several waves as they evolve.

\smallskip

To provide further clarification of the model, we mention that the skew-bi logistic distribution is {\em not} a {\em mixture model} per se, in which case there is a mixture weight for each distribution in the sum, as in, say, a Gaussian mixture \cite[Chapter 9]{BISH06}. In the bi-skew logistic distribution case we do not have mixture weights, rather we have two phases or in our context waves, which are sequential in nature, possibly with some overlap, as can be seen in Figure~\ref{fig:deaths} (c.f. \cite{MEYE94,SHEE04}); however, strictly speaking, the bi-skew logistic distribution can be viewed as a mixture model where the mixture weights are each $0.5$ and a scaling factor of $2$ is applied. Thus, as an approximation, we add a preprocessing step where we segment the time series into distinct waves, resulting in a considerable reduction to the complexity of the maximum likelihood estimation. We do, however, remark that the maximum likelihood estimation for the bi-skew logistic distribution is much simpler than that of a corresponding mixture model, due to the absence of mixture weights. In particular, although we could, in principle make use of the EM (expectation-maximisation) algorithm \cite{REDN84} \cite[Chapter 9]{BISH06} to approximate the maximum likelihood estimates of the parameters, this would not be strictly necessary in the bi-skew logistic case, cf. \cite{MCDO21}. The only caveat, which holds independently of whether the EM algorithm is deployed or not, is the additional number of parameters present in the equations being solved. We leave this investigation as future work, and focus on our approximation, which does not require the solution to the maximum likelihood of (\ref{eq:bi-skew}); the details of the preprocessing heuristic we apply are given in the following section.

\section{Data analysis of COVID-19 deaths in the UK}
\label{sec:data}

Here we provide a full analysis of COVID-19 deaths in the UK from 30/01/20 to 30/07/21, employing the ${\cal E}SJS$ goodness-of-fit statistic and comparing it to the $KS2$ statistic. The daily UK COVID-19 data we used was obtained from \cite{GOV21}.

\smallskip

As a proof of concept of the modelling capability of the skew logistic distribution, we now provide a detailed analysis of the time series of COVID-19 deaths in the UK from 30/01/20 to 30/07/21.

\smallskip

To separate the waves we first smoothed the raw data using a moving average with a centred sliding window of 7 days.
We then applied a simple heuristic, where we identified all the minima in the time series and defined a wave as a consecutive portion of the time, of at least 72 days, with the endpoints of each wave being local minima apart from the first wave which starts from day 0. The resulting four waves in the time series are shown in Figure~\ref{fig:deaths}; see last column of Table~\ref{table:sl} for the endpoints of the four waves. It would be worthwhile, as future work, to investigate other heuristics, which may for example allow overlap between the waves to obtain more accurate start and end points and to distribute the number of cases between the waves when there is overlap between them.

\begin{figure}[ht]
\begin{center}
\includegraphics[scale=0.6]{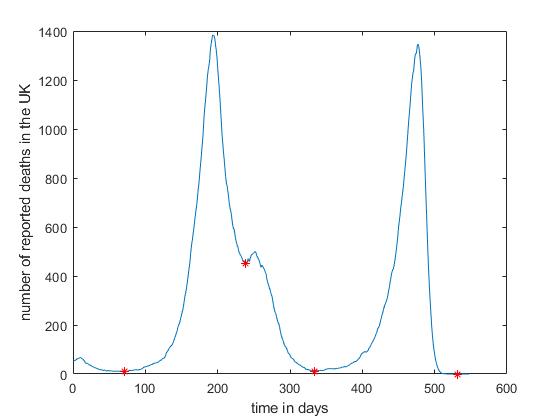}
\caption{\label{fig:deaths} Reported daily COVID-19 deaths from 30/01/20 to 30/07/21 and their minima labelled `*', resulting in four distinct waves; a moving average with a centred sliding window of 7 days was applied to the raw data.}
\end{center}
\end{figure}
\smallskip

In Table~\ref{table:sl} we show the parameters resulting from maximum likelihood fits of the skew logistic distribution to the four waves. Figure~\ref{fig:waves} shows histograms of the four COVID-19 waves, each overlaid with the curve of the maximum likelihood fit of the skew logistic distribution to the data. Pearson's moment and median skewness coefficients \cite{DOAN11} for the four waves are recorded in Table~\ref{table:skew}. It can be seen that the correlation between these and $1-\lambda$ is close to $1$, as we would expect.

\begin{table}[ht]
\begin{center}
\begin{tabular}{|l|c|c|c|c|}\hline
\multicolumn{5}{|c|}{Fitted parameters for the skew logistic distribution} \\ \hline
Wave & $\lambda$ & $\mu$    & $s$     & End \\ \hline \hline
1    & 0.2150    & 3.5137   & 3.8443  & 71  \\ \hline
2    & 1.0741    & 196.5157 & 14.4323 & 239 \\ \hline
3    & 0.2297    & 243.0709 & 4.5882  & 334 \\ \hline
4    & 1.7306    & 502.2758 & 7.0195  & 532 \\ \hline
\end{tabular}
\end{center}
\caption{\label{table:sl} Parameters from maximum likelihood fits of the skew logistic distribution to the four waves, and the day of the local minimum (End), which is the end point of the wave.}
\end{table}

\begin{figure}[ht]
\begin{center}
\includegraphics[scale=0.75]{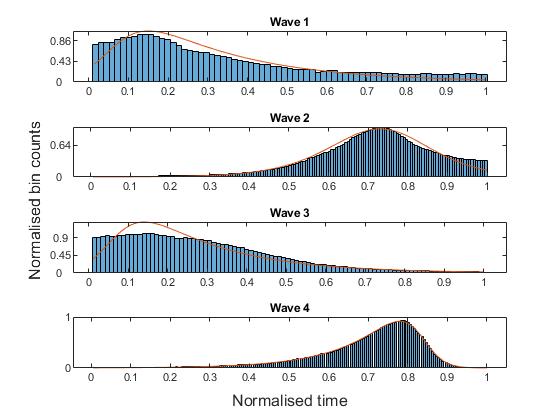}
\caption{\label{fig:waves} Histograms for the four waves of COVID-19 deaths from 30/01/20 to 30/07/21, each overlaid with
the curve of the maximum likelihood fit of the skew logistic distribution to the data.}
\end{center}
\end{figure}

\begin{table}[ht]
\begin{center}
\begin{tabular}{|l|c|c|c|}\hline
\multicolumn{4}{|c|}{Skewness} \\ \hline
Wave & $1-\lambda$ & moment  & median                \\ \hline \hline
1    & 0.7850      &  0.9314 &  0.2939               \\ \hline
2    & -0.0741     & -0.7758 & -0.0797               \\ \hline
3    & 0.7703      &  0.9265 &  0.1939               \\ \hline
4    & -0.7306     & -1.5555 & -0.2413               \\ \hline
 \multicolumn{2}{|c|}{Correlation} & 0.9931 & 0.9826 \\ \hline
\end{tabular}
\end{center}
\caption{\label{table:skew} Pearson's moment and median skewness coefficients for the four waves, and the correlation between $1-\lambda$ and these coefficients.}
\end{table}
\smallskip

We now turn to the evaluation of goodness-of-fit using the ${\cal E}SJS$ (empirical survival Jensen-Shannon divergence) \cite{LEVE18,LEVE21a}, which generalises the Jensen-Shannon divergence \cite{LIN91} to survival functions, and the well-known $KS2$ (Kolmogorov-Smirnov two-sample test statistic) \cite[Section 6.3]{GIBB21}. We will also employ 95\% bootstrap confidence intervals \cite{EFRO93} to measure the improvement in the ${\cal E}SJS$ and $KS2$, goodness-of-fit measures, of the  skew-logistic over the logistic and normal distributions, respectively. For completeness we formally define the ${\cal E}SJS$ and $KS2$.

\smallskip

To set the scene we assume a time series \cite{CHAT19}, ${\bf x} = \{x_1, x_2, \ldots, x_n\}$, where $x_t$, for $t=1,2,\ldots, n$ is a value indexed by time, $t$, in our case modelling the number of daily COVID-19 deaths. We are, in particular, interested in the marginal distribution of ${\bf x}$, which we suppose comes from an underlying parametric continuous distribution $D$.

\smallskip

The {\em empirical survival function} of a value $z$ for the time series ${\bf x}$, denoted by $\widehat{S}({\bf x})[z]$, is given by
\begin{equation}\label{eq:emp}
\widehat{S}({\bf x})[z] = \frac{1}{n} \sum_{i=1}^n I_{\{x_i> z\}},
\end{equation}
where $I$ is the indicator function. In the following we will let $\widehat{P}(z) = \widehat{S}({\bf x})[z]$  stand for the empirical survival function $\widehat{S}({\bf x})[z]$, where the time series ${\bf x}$ is assumed to be understood from context; we will generally be interested in the empirical survival function $\widehat{P}$, which we suppose arises from the survival function $P$ of the parametric continuous distribution $D$, mentioned above.

\smallskip

The {\em empirical survival Jensen-Shannon divergence} (${\cal E}SJS$) between two empirical survival functions, $\widehat{Q}_1$ and $\widehat{Q}_2$ arising from the survival functions $Q_1$ and $Q_2$, is given by
\begin{equation}\label{eq:ejs}
{\cal E}SJS(\widehat{Q}_1,\widehat{Q}_2) = \frac{1}{2} \ \int_0^\infty \ \widehat{Q}_1(z) \ \log \left( \frac{\widehat{Q}_1(z)}{\widehat{M}(z} \right) \ + \ \widehat{Q}_2(z) \ \log \left( \frac{\widehat{Q}_2(z)}{\widehat{M}(z)} \right) {\rm d} z,
\end{equation}
where
\begin{displaymath}
\widehat{M}(z) = \frac{1}{2} \ \left( \widehat{Q}_1(z) \ + \ \widehat{Q}_2(z) \right).
\end{displaymath}
\smallskip

We note that the ${\cal E}SJS$ is bounded and can thus be normalised, so it is natural to assume its values are between $0$ and $1$; in particular, when $\widehat{Q}_1 = \widehat{Q}_2$ its value is zero. Moreover, its square root is a metric \cite{NGUY15}, cf. \cite{LEVE18}.

\smallskip

The {\em Kolmogorov-Smirnov} two-sample test statistic between $\widehat{Q}_1$ and $\widehat{Q}_2$ as above, is given by
\begin{equation}\label{eq:ks2}
KS2(\widehat{Q}_1, \widehat{Q}_2) = \underset{z}{max} | \widehat{Q}_1(z) - \widehat{Q}_2(z) |,
\end{equation}
where $max$ is the maximum function, and $|v|$ is the absolute value of a number $v$.
We note that $KS2$ is bounded between $0$ and $1$, and is also a metric.

\smallskip

For a parametric continuous distribution $D$, we let $\phi = \phi(D,\widehat{P})$ be the parameters that are obtained from fitting $D$ to the empirical survival function, $\widehat{P}$, using maximum likelihood estimation. In addition, we let $P_\phi = S_\phi({\bf x})$ be the survival function of ${\bf x}$, for $D$ with parameters $\phi$.
Thus, the empirical survival Jensen-Shannon divergence and the Kolmogorov-Smirnov two-sample test statistic, between $\widehat{P}$ and $P_\phi$, are given by ${\cal E}SJS(\widehat{P}, P_\phi)$ and $KS2(\widehat{P},P_\phi)$, respectively, where $\widehat{P}$ and $P_\phi$ are omitted below as they will be understood from context. These values provide us with two measures of goodness-of-fit for how well $D$ with parameters $\phi$ is fitted to ${\bf x}$ \cite{LEVE21a}.

\smallskip

We are now ready to present the results of the evaluation.
In Table~\ref{table:esjs} we show the ${\cal E}SJS$ values for the four waves and the said improvements, while in  Table~\ref{table:ks2} we show the corresponding $KS2$ values and improvements. In all cases the skew logistic is a preferred model over both the logistic and normal distributions, justifying the addition of a skewness parameter as can be see in
Figure~\ref{fig:waves}. Moreover, in all but one case is the logistic distribution preferred over the normal distribution; this is for wave 3, where the $KS2$ statistic of the normal distribution is smaller than that of the logistic distribution. We observe that, for the second wave, the ${\cal E}SJS$ and $KS2$ values for the skew logistic and logistic distribution are the closest, since as can be seen from Table~\ref{table:sl} the second wave was more or less symmetric, in which case the skew logistic distribution reduces to the logistic distribution.

\begin{table}[ht]
\begin{center}
\begin{tabular}{|l|c|c|c|c|c|}\hline
\multicolumn{6}{|c|}{${\cal E}SJS$ values for SL, Logit and Norm distributions} \\ \hline
Wave & SL     & Logit  & SL-Logit & Norm   & SL-Norm \\ \hline \hline
1    & 0.0419 & 0.0583 & 28.25\%  & 0.0649 & 35.54\% \\ \hline
2    & 0.0392 & 0.0448 & 12.52\%  & 0.0613 & 36.17\% \\ \hline
3    & 0.0316 & 0.0387 & 18.38\%  & 0.0423 & 25.38\% \\ \hline
4    & 0.0237 & 0.0927 & 74.47\%  & 0.0939 & 74.79\% \\ \hline
\end{tabular}
\end{center}
\caption{\label{table:esjs} ${\cal E}SJS$ values for the skew logistic (SL), logistic (Logit) and normal (Norm) distributions, and the improvement percentage of the skew logistic over the logistic (SL-Logit) and normal (SL-Norm) distributions, respectively.}
\end{table}

\begin{table}[ht]
\begin{center}
\begin{tabular}{|l|c|c|c|c|c|}\hline
\multicolumn{6}{|c|}{$KS2$ values for SL, Logit and Norm distributions} \\ \hline
Wave & SL     & Logit  & SL-Logit & Norm   & SL-Norm \\ \hline \hline
1    & 0.0621 & 0.1245 & 50.14\%  & 0.1280 & 51.50\% \\ \hline
2    & 0.0357 & 0.0391 & 8.57\%   & 0.0420 & 15.01\% \\ \hline
3    & 0.0571 & 0.0930 & 38.66\%  & 0.0854 & 33.18\% \\ \hline
4    & 0.0098 & 0.0817 & 87.98\%  & 0.1046 & 90.61\% \\ \hline
\end{tabular}
\end{center}
\caption{\label{table:ks2} $KS2$ values for the skew logistic (SL), logistic (Logit) and normal (Norm) distributions, and the improvement percentage of the skew logistic over the logistic (SL-Logit) and normal (SL-Norm) distributions, respectively.}
\end{table}
\smallskip

In Tables \ref{table:boot-jsd} and \ref{table:boot-ks2} we present the bootstrap 95\% confidence intervals of the ${\cal E}SJS$ and $KS2$ improvements, respectively, using the {\em percentile} method, while in Tables \ref{table:bca-jsd} and \ref{table:bca-ks2} we provide the 95\% confidence intervals of the ${\cal E}SJS$ and $KS2$ improvements, respectively, using the {\em bias-corrected and accelerated} (BCa) method \cite{EFRO93}, which adjusts the confidence intervals for bias and skewness in the empirical bootstrap distribution. In all cases the mean of the  bootstrap samples is above zero with a very tight standard deviation. As noted above the second wave is more or less symmetric, so we expect that the standard logistic distribution will provide a fit to the data which as good as the skew logistic fit. It is thus not surprising that in this case the improvement percentages are, generally, not significant. In addition, the improvements for the third wave are also, generally, not significant, which may be due to the starting point of the third wave, given our heuristic, being close to its peak; see Figure~\ref{fig:deaths}. We observe that, for this data set, it is not clear whether deploying the BCa method yields a significant advantage over simply deploying of the percentile method.

\smallskip

In Table~\ref{table:mean-std} we show the mean and standard deviation statistics of the confidence interval widths, of the metrics we used to compare the distributions, implying that, in general, the ${\cal E}SJS$ goodness-of-fit measure is more powerful than the $KS2$ goodness-of-fit measure. This is based on the known result that statistical tests using measures resulting in smaller confidence intervals are normally considered to be more powerful, implying that a smaller sample size may be deployed  \cite{LIU13a}.

\begin{table}[ht]
\begin{center}
\begin{tabular}{|l|c|c|c|c|c|}\hline
\multicolumn{6}{|c|}{Percentile confidence intervals for ${\cal E}SJS$ improvement} \\ \hline
Wave/Diff  & LB of CI      & UB of CI & Width of CI & Mean   & STD    \\ \hline \hline
1/SL-Logit & 0.0093        & 0.0317   &  0.0224     & 0.0211 & 0.0063 \\ \hline
1/SL-Norm  & 0.0170        & 0.0382   &  0.0212     & 0.0278 & 0.0063 \\ \hline
2/SL-Logit & {\em -0.0010} & 0.0066   &  0.0076     & 0.0034 & 0.0049 \\ \hline
2/SL-Norm  & 0.0154        & 0.0232   &  0.0078     & 0.0201 & 0.0051 \\ \hline
3/SL-Logit & {\em -0.0028} & 0.0112   &  0.0140     & 0.0083 & 0.0022 \\ \hline
3/SL-Norm  & 0.0021        & 0.0149   &  0.0128     & 0.0120 & 0.0022 \\ \hline
4/SL-Logit & 0.0549        & 0.0810   &  0.0261     & 0.0714 & 0.0068 \\ \hline
4/SL-Norm  & 0.0560        & 0.0821   &  0.0261     & 0.0722 & 0.0070 \\ \hline
\end{tabular}
\end{center}
\caption{\label{table:boot-jsd} Results from the percentile method for the confidence interval of the difference of the ${\cal E}SJS$ between the logistic (Logit) and skew logistic (SL), and between the normal (Norm) and skew logistic (SL) distributions, respectively; Diff, LB, UB, CI, Mean and STD stand for difference, lower bound, upper bound, confidence interval, mean of samples and standard deviation of samples, respectively.}
\end{table}

\begin{table}[ht]
\begin{center}
\begin{tabular}{|l|c|c|c|c|c|}\hline
\multicolumn{6}{|c|}{Percentile confidence intervals for $KS2$ improvement} \\ \hline
Wave/Diff  & LB of CI      & UB of CI & Width of CI & Mean   & STD  \\ \hline \hline
1/SL-Logit & 0.0438        & 0.0760   & 0.0322      & 0.0621 & 0.0073 \\ \hline
1/SL-Norm  & 0.0411        & 0.0821   & 0.0410      & 0.0684 & 0.0078 \\ \hline
2/SL-Logit & 0.0003        & 0.0047   & 0.0044      & 0.0033 & 0.0009\\ \hline
2/SL-Norm  & 0.0007        & 0.0092   & 0.0085      & 0.0065 & 0.0017 \\ \hline
3/SL-Logit & {\em -0.0073} & 0.0441   & 0.0514      & 0.0343 & 0.0082\\ \hline
3/SL-Norm  & {\em -0.0142} & 0.0365   & 0.0507      & 0.0267 & 0.0080   \\ \hline
4/SL-Logit & 0.0474        & 0.0728   & 0.0254      & 0.0680 & 0.0046\\ \hline
4/SL-Norm  & 0.0710        & 0.0962   & 0.0252      & 0.0905 & 0.0048 \\ \hline
\end{tabular}
\end{center}
\caption{\label{table:boot-ks2} Results from the percentile method for the confidence interval of the difference of the $KS2$ between the logistic (Logit) and skew logistic (SL), and between the normal (Norm) and skew logistic (SL) distributions, respectively; Diff, LB, UB, CI, Mean and STD stand for difference, lower bound, upper bound, confidence interval, mean of samples and standard deviation of samples, respectively.}
\end{table}

\begin{table}[ht]
\begin{center}
\begin{tabular}{|l|c|c|c|c|c|}\hline
\multicolumn{6}{|c|}{BCa confidence intervals for ${\cal E}SJS$ improvement} \\ \hline
Wave/Diff  & LB of CI       & UB of CI & Width of CI & Mean   & STD    \\ \hline \hline
1/SL-Logit & 0.0087         & 0.0260   & 0.0173      & 0.0210 & 0.0062 \\ \hline
1/SL-Norm  & 0.0165         & 0.0333   & 0.0168      & 0.0275 & 0.0063 \\ \hline
2/SL-Logit & {\em  -0.0009} & 0.0258   & 0.0267      & 0.0036 & 0.0053 \\ \hline
2/SL-Norm  & 0.0153         & 0.0425   & 0.0272      & 0.0201 & 0.0050 \\ \hline
3/SL-Logit & {\em -0.0024}  & 0.0095   & 0.0119      & 0.0084 & 0.0023 \\ \hline
3/SL-Norm  & {\em -0.0027}  & 0.0135   & 0.0162      & 0.0119 & 0.0024 \\ \hline
4/SL-Logit & 0.0308         & 0.0703   & 0.0395      & 0.0708 & 0.0074 \\ \hline
4/SL-Norm  & 0.0554         & 0.0713   & 0.0159      & 0.0726 & 0.0069 \\ \hline
\end{tabular}
\end{center}
\caption{\label{table:bca-jsd} Results from the BCa method for the confidence interval of the difference of the ${\cal E}SJS$ between the logistic (Logit) and skew logistic (SL), and between the normal (Norm) and skew logistic (SL) distributions, respectively;  Diff, LB, UB, CI, Mean and STD stand for difference, lower bound, upper bound, confidence interval, mean of samples and standard deviation of samples, respectively.}
\end{table}

\begin{table}[ht]
\begin{center}
\begin{tabular}{|l|c|c|c|c|c|}\hline
\multicolumn{6}{|c|}{BCa confidence intervals for $KS2$ improvement} \\ \hline
Wave/Diff  & LB of CI      & UB of CI & Width of CI & Mean   & STD  \\ \hline \hline
1/SL-Logit & 0.0428        & 0.0801   & 0.0373      & 0.0624 & 0.0074 \\ \hline
1/SL-Norm  & 0.0444        & 0.0777   & 0.0333      & 0.0683 & 0.0078 \\ \hline
2/SL-Logit & 0.0005        & 0.0047   & 0.0042      & 0.0033 & 0.0008 \\ \hline
2/SL-Norm  & 0.0001        & 0.0089   & 0.0088      & 0.0064 & 0.0017 \\ \hline
3/SL-Logit & 0.0013        & 0.0445   & 0.0432      & 0.0346 & 0.0077 \\ \hline
3/SL-Norm  & {\em -0.0111} & 0.0368   & 0.0479      & 0.0263 & 0.0082 \\ \hline
4/SL-Logit & 0.0491        & 0.0739   & 0.0248      & 0.0676 & 0.0047 \\ \hline
4/SL-Norm  & 0.0685        & 0.0985   & 0.0300      & 0.0908 & 0.0046  \\ \hline
\end{tabular}
\end{center}
\caption{\label{table:bca-ks2} Results from the BCa method for the confidence interval of the difference of the $KS2$ between the logistic (Logit) and skew logistic (SL), and between the normal (Norm) and skew logistic (SL) distributions, respectively;  Diff, LB, UB, CI, Mean and STD stand for difference, lower bound, upper bound, confidence interval, mean of samples and standard deviation of samples, respectively.}
\end{table}

\begin{table}[ht]
\begin{center}
\begin{tabular}{|l|c|c|c|c|}\hline
\multicolumn{5}{|c|}{Summary statistics for the CI widths} \\ \hline
Statistic & ${\cal E}SJS$-P & $KS2$-P & ${\cal E}SJS$-BCa & $KS2$-BCa \\ \hline \hline
Mean      & 0.0172 & 0.0298 & 0.0214 & 0.0287 \\ \hline
STD       & 0.0077 & 0.0176 & 0.0091 & 0.0155 \\ \hline
\end{tabular}
\end{center}
\caption{\label{table:mean-std} Mean and standard deviation (STD) statistics for the confidence interval (CI) widths using the percentile (P) and BCa methods.}
\end{table}
\smallskip

As mentioned in the introduction, we obtained comparable results to the above when modelling epidemic waves with the epsilon skew normal distribution \cite{MUDH00} as opposed to using the skew logistic distribution; see also \cite{KAZE15} for a comparison of a skew logistic and skew normal distributions in the context of insurance loss data, showing that the skew logistic performed better than the skew normal distribution for fitting the data sets tested. Further to the note in the introduction, that the skew logistic distribution is a more natural one to deploy in this case due to its heavier tails, we observe that in an epidemic scenario the number of cases counted can only be non-negative, while the epsilon skew normal also supports negative values.

\section{Concluding remarks}
\label{sec:conc}

We have proposed the skew-logistic and bi-logistic distributions as models for single and multiple epidemic waves, respectively. The model is a simple extension of the symmetric logistic distribution, which can readily be deployed in the presence of skewed data that exhibits growth and decay. We provided validation for the proposed model using the ${\cal E}SJS$ as a goodness-of-fit statistic, showing that it is a good fit to COVID-19 data in UK and more powerful than the alternative $KS2$ statistic. As future work, we could use the model to compare the progression of multiple waves across different countries, extending the work of \cite{DYE20}.

\newcommand{\etalchar}[1]{$^{#1}$}


\begin{thebibliography}{{GOV}21}

\bibitem[AC14]{AZZA14}
A.~Azzalini and A.~Capitanio.
\newblock {\em The Skew-Normal and Related Families}.
\newblock Institute Of Mathematical Statistics Monographs. Cambridge University
  Press, Cambridge, UK, 2014.

\bibitem[AS72]{ABRA72}
M.~Abramowitz and I.A. Stegun, editors.
\newblock {\em Handbook of Mathematical Functions with Formulas, Graphs and
  Mathematical Tables}.
\newblock Dover, New York, NY, 1972.

\bibitem[Bac11]{BACA11}
N.~Baca\"{e}r.
\newblock {\em A Short History of Mathematical Population Dynamics}.
\newblock Springer Verlag, London, 2011.

\bibitem[Bis06]{BISH06}
C.M. Bishop.
\newblock {\em Pattern Recognition and Machine Learning}.
\newblock Information Science and Statistics. Springer Science+Business Media,
  New York, NY, 2006.

\bibitem[BLJ15]{BEND15}
A.~{Ben-David}, H.~Liu, and A.D. Jackson.
\newblock The {Kullback-Leibler} divergence as an estimator of the statistical
  properties of {CMB} maps.
\newblock {\em Journal of Cosmology and Astroparticle Physics}, 2015:JCAP06051,
  June 2015.

\bibitem[CH82]{CLIFF82}
A.~Cliff and P.~Haggett.
\newblock Methods for the measurement of epidemic velocity from time-series
  data.
\newblock {\em International Journal of Epidemiology}, 11:82--89, 1982.

\bibitem[CR19]{COLE21}
N.~Colegrave and G.D. Ruxton.
\newblock {\em Power Analysis: An introduction for the life sciences}.
\newblock Oxford Biology Primers. Oxford University Press, Oxford, 2019.

\bibitem[CX19]{CHAT19}
C.~Chatfield and H.~Xing.
\newblock {\em The Analysis of Time Series: An Introduction with R}.
\newblock Text in Statistical Science. Chapman \& Hall, London, 7th edition,
  2019.

\bibitem[Das10]{DASG10}
A.~DasGupta.
\newblock {\em Fundamentals of Probability: A First Course}.
\newblock Springer Texts in Statistics. Springer Science+Business Media, New
  York, NY, 2010.

\bibitem[DCDW20]{DYE20}
C.~Dye, R.C.H. Cheng, J.S. Dagpunar, and B.G. Williams.
\newblock The scale and dynamics of {COVID}-19 epidemics across europe.
\newblock {\em Royal Society Open Science}, 7:201726--1--201726--8, 2020.

\bibitem[DI20]{DELA20}
M.~{De la Sen} and A.~Ibeas.
\newblock On an {S}ir epidemic model for the {COVID-19} pandemic and the
  logistic equation.
\newblock {\em Discrete Dynamics in Nature and Society}, Article ID 1382870:17
  pages, 2020.

\bibitem[DKE{\etalchar{+}}20]{DAVI20}
N.G. Davies, A.J. Kucharski, R.M. Eggo, A.Gimma, and W.J. Edmunds.
\newblock Effects of non-pharmaceutical interventions on {COVID}-19 cases,
  deaths, and demand for hospital services in the {UK}: {A} modelling study.
\newblock {\em THE LANCET Public Health}, 5:e375--e385, 2020.

\bibitem[DS11]{DOAN11}
D.P. Doane and L.E. Seward.
\newblock Measuring skewness {A} forgotten statistic?
\newblock {\em Journal of Statistics Education}, 19(2):19 pages, 2011.

\bibitem[ET93]{EFRO93}
B.~Efron and R.~Tibshirani.
\newblock {\em An Introduction to the Bootstrap}.
\newblock Monographs on Statistics and Applied Probability 57. Springer
  Science+Business Media, New York, NY, 1993.

\bibitem[FLL13]{FENN13}
T.~Fenner, M.~Levene, and G.~Loizou.
\newblock A bi-logistic growth model for conference registration with an early
  bird deadline.
\newblock {\em Central European Journal of Physics}, 11:904--909, 2013.

\bibitem[GC21]{GIBB21}
J.D. Gibbons and S.~Chakraborti.
\newblock {\em Nonparametric Statistical Inference}.
\newblock Marcel Dekker, New York, NY, sixth edition, 2021.

\bibitem[{GOV}21]{GOV21}
{GOV.UK}.
\newblock {C}oronavirus ({COVID}-19) in the {UK}, {D}ownload data.
\newblock See \url{https://coronavirus.data.gov.uk/details/download}, 2021.

\bibitem[HKT21]{HARV21}
A.~Harvey, P.~Kattuman, and C.~Thamotheram.
\newblock Tracking the mutant: {F}orecasting and nowcasting {COVID}-19 in the
  {UK} in 2021.
\newblock {\em National Institute Economic Review}, 256:110--126, 2021.

\bibitem[ICT22]{IOAN22}
J.P.A. Ioannidis, S.~Cripps, and M.A. Tanner.
\newblock Forecasting for {COVID}-19 has failed.
\newblock {\em International Journal of Forecasting}, 38:423--438, 2022.

\bibitem[JKB95]{JOHN95c}
N.L. Johnson, S.~Kotz, and N.~Balkrishnan.
\newblock {\em Continuous Univariate Distributions, Volume 2}, chapter 23
  Logistic distribution, pages 113--163.
\newblock Wiley Series in Probability and Mathematical Statistics. John Wiley
  {\&} Sons, New York, NY, second edition, 1995.

\bibitem[KN15]{KAZE15}
R.~Kazemi and M.~Noorizadeh.
\newblock A comparison between skew-logistic and skew-normal distributions.
\newblock {\em Matematika}, 31:15--24, 2015.

\bibitem[Kri15]{KRIS15}
K.~Krishnamoorthy.
\newblock {\em Handbook of Statistical Distributions with Applications}.
\newblock CRC Press, Boca Raton, FL, second edition, 2015.

\bibitem[Lev21]{LEVE21a}
M.~Levene.
\newblock A hypothesis test for the goodness-of-fit of the marginal
  distribution of a time series with application to stablecoin data.
\newblock {\em Engineering Proceedings}, 5:Article 10, 8 pages, 2021.
\newblock Presented at the International conference on Time Series and
  Forecasting (ITISE).

\bibitem[Li18]{LI18}
M.Y. Li.
\newblock {\em An Introduction to Mathematical Modeling of Infectious
  Diseases}.
\newblock Mathematics of Planet Earth. Springer Nature, Cham, Switzerland,
  2018.

\bibitem[Lin91]{LIN91}
J.~Lin.
\newblock Divergence measures based on the {S}hannon entropy.
\newblock {\em IEEE Transactions on Information Theory}, 37:145--151, 1991.

\bibitem[Liu13]{LIU13a}
X.S. Liu.
\newblock Comparing sample size requirements for significance tests and
  confidence intervals.
\newblock {\em Counseling Outcome Research and Evaluation}, 4:3--12, 2013.

\bibitem[LK21]{LEVE18}
M.~Levene and A.~Kononovicius.
\newblock Empirical survival {Jensen-Shannon} divergence as a goodness-of-fit
  measure for maximum likelihood estimation and curve fitting.
\newblock {\em Communications in Statistics - Simulation and Computation},
  50:3751--3767, 2021.

\bibitem[Mac21]{MCDO21}
I.L. Mac{D}onald.
\newblock Is {EM} really necessary here? {E}xamples where it seems simpler not
  to use {EM}.
\newblock {\em AStA Advances in Statistical Analysis}, 105:629--–647, 2021.

\bibitem[Mey94]{MEYE94}
P.~Meyer.
\newblock Bi-logistic growth.
\newblock {\em Technological Forecasting and Social Change}, 47:89--102, 1994.

\bibitem[MH00]{MUDH00}
G.S. Mudholkar and A.D. Hutson.
\newblock The epsilon--skew--normal distribution for analyzing near-normal
  data.
\newblock {\em Journal of Statistical Planning and Inference}, 83:291--309,
  2000.

\bibitem[Nad09]{NADA09a}
S.~Nadarajah.
\newblock The skew logistic distribution.
\newblock {\em AStA Advances in Statistical Analysis}, 93:187--–203, 2009.

\bibitem[NV15]{NGUY15}
{H.-V.} Nguyen and J.~Vreeken.
\newblock Non-parametric {Jensen-Shannon} divergence.
\newblock In {\em Proceedings of European Conference on Machine Learning and
  Principles and Practice of Knowledge Discovery in Databases (ECML PKDD)},
  pages 173--189, Porto, 2015.

\bibitem[Pan14]{PANI14}
M.J. Panik.
\newblock {\em Growth Curve Modeling: Theory and Applications}.
\newblock John Wiley {\&} Sons, Hoboken, NJ, 2014.

\bibitem[PKK{\etalchar{+}}20]{PELIN20}
E.~Pelinovsky, A.~Kurkin, O.~Kurkina, M.~Kokoulina, and A.~Epifanova.
\newblock Logistic equation and {COVID-19}.
\newblock {\em Chaos, Solitons and Fractals}, 140:110241--1--110241--13, 2020.

\bibitem[POP17]{PEAR17}
J.W. Pearson, S.~Olver, and M.A. Porter.
\newblock Numerical methods for the computation of the confluent and {G}auss
  hypergeometric functions.
\newblock {\em Numerical Algorithms}, 74:821--866, 2017.

\bibitem[Pos20]{POST20}
E.B. Postnikov.
\newblock Estimation of {COVID-19} dynamics “on a back-of-envelope”: {D}oes
  the simplest {SIR} model provide quantitative parameters and predictions?
\newblock {\em Chaos, Solitons and Fractals}, 135:109841--1--109841--6, 2020.

\bibitem[RW84]{REDN84}
R.A. Redner and H.F. Walker.
\newblock Mixture densities, maximum likelihood and the {Em} algorithm.
\newblock {\em SIAM Review}, 26:195--239, 1984.

\bibitem[SB16]{SAST16}
D.V.S. Sastry and D.~Bhati.
\newblock A new skew logistic distribution: {P}roperties and applications.
\newblock {\em Brazilian Journal of Probability and Statistics}, 30:248--271,
  2016.

\bibitem[SJ22]{SEME22}
I.~Semenov and M.~Jacyna.
\newblock The synthesis model as a planning tool for effective supply chains
  resistant to adverse events.
\newblock {\em Maintenance and Reliability}, 24:140--152, 2022.

\bibitem[SMF04]{SHEE04}
J.E. Sheehy, P.L. Mitchell, and A.B. Ferrer.
\newblock Bi-phasic growth patterns in rice.
\newblock {\em Annals of Botany}, 94:811--817, 2004.

\bibitem[SN10]{SPIE10}
A.-N. Spiess and N.~Neumeyer.
\newblock An evaluation of {$R^2$} as an inadequate measure for nonlinear
  models in pharmacological and biochemical research: a {Monte Carlo} approach.
\newblock {\em BMC Pharmacology}, 10, 2010.
\newblock 11 pages.

\bibitem[WA18]{WARD18}
M.D. Ward and J.S. Ahlquist.
\newblock {\em Maximum Likelihood for Social Science: Strategies for Analysis}.
\newblock Analytical Methods for Social Research. Cambridge University Press,
  Cambridge, UK, 2018.

\bibitem[ZMG21]{ZHAN21}
S.X. Zhang, F.A. Marioli, and R.~Gao.
\newblock A second wave? {W}hat do people mean by {COVID} waves? {\textendash}
  a working definition of epidemic waves.
\newblock {\em Risk Management and Healthcare Policy}, 14:3775--3782, 2021.

\end{thebibliography}
\end{document}